\documentclass[a4paper]{jpconf}
\usepackage{graphicx}
\usepackage{amssymb,multirow,epstopdf,xcolor,hyperref}

\graphicspath{{fig/}}

\usepackage[normalem]{ulem}
\usepackage{color}
\definecolor{linkcolor}{HTML}{799B03}
\definecolor{urlcolor}{HTML}{799B03}

\hypersetup{linkcolor=linkcolor, urlcolor=urlcolor, colorlinks=true}

\begin{document}
\title{Helicity of convective flows from localized heat source in a rotating layer}

\author[icmm]{A. Sukhanovskii, A. Evgrafova, E. Popova}

\address{Institute of Continuous Media Mechanics -  Academ. Korolyov, 1, Perm, 614013, Russia}

\ead{san@icmm.ru}

\begin{abstract}
Experimental and numerical study of the steady-state cyclonic vortex from isolated heat source in a rotating fluid layer is described. The structure of laboratory cyclonic vortex is similar to the typical structure of tropical cyclones from observational data and numerical modelling including secondary flows in the boundary layer. Differential characteristics of the flow were studied by numerical simulation using CFD software FlowVision. Helicity distribution in rotating fluid layer with localized heat source was analysed. Two mechanisms which play role in helicity generation are found. The first one is the strong correlation of cyclonic vortex and intensive upward motion in the central part of the vessel. The second one is due to large gradients of velocity on the periphery. The integral helicity in the considered case is substantial and its relative level is high.
\end{abstract}

\section{Introduction}
Despite decades of research, the problem of tropical cyclogenesis is unsolved and attracts close attention from many scientific groups.  The complexity of the problem forces researchers to study tropical cyclogenesis step by step seeing as a main goal the theory that would describe all stages of the tropical cyclone formation. Laboratory model of hurricane-like vortex was proposed and studied in [1-3], where rotating layer of fluid with the localized heater in the bottom was considered. The important differences of this experimental approach were more viscous fluids (in comparison with water) and the use of a shallow layer. Later, series of the experiments [4] were done for the same configuration using PIV system for velocity measurements. The main focus of [4] was on integral characteristics of the azimuthal flows such as angular momentum and kinetic energy. Detailed study of different constraints of the steady-state hurricane-like vortex were studied in [5]. The three main dimensional parameters that define the vortex structure for a fixed geometry -- heating flux, rotation rate and viscosity were varied independently. It was shown that viscosity is one of the main parameters that define steady-state vortex structure. Increasing of kinematic viscosity may substantially suppress the cyclonic motion for fixed values of buoyancy flux and rotation rate. Strong competition between buoyancy and rotation provides the optimal ratio of the heating flux and rotation rate for achieving cyclonic vortex of maximal intensity. It was found that relatively small variation of the rotation rate for the fluids with low kinematic viscosity may remarkably change the cyclonic vortex structure and  intensity.

Here we focused our attention on differential characteristics of the convective flow, mainly helicity which is a scalar production of velocity ($\upsilon$) and vorticity ($\nabla \times \textbf{$\upsilon$}$) vectors. The volume integral
\begin{equation}
H = \int \textbf{$\upsilon$} \cdot (\nabla \times \textbf{$\upsilon$})dV = \int\limits_0^{2\pi}\int\limits_0^r\int\limits_0^z h d\phi dr dz=\int\limits_0^r\int\limits_0^z 2\pi r\cdot h dr dz
\label{hel}
\end{equation}
gives the total (or global) helicity of vortex system, where $h=\textbf{$\upsilon$} \cdot \nabla \times \textbf{$\upsilon$}$ is the helicity density of the flow. In early eighties, in a series of papers [6-9], the role of helicity in formation and dynamics of intensive vortices was discussed. Up to now this problem is unsolved. Helicity is an invariant such as energy or angular momentum and its generation or conservation may change energy cascade in a developed turbulent media. The problem of helicity of turbulent flows is very complex for studying, because it requires measurements of 3D velocity field in a volume or high-resolution numerical modelling. Even realization of the flow with substantial value of helicity is a complicated problem [10]. Here, we consider helicity in a laboratory hurricane-like vortex. The flow in our experimental system is not fully turbulent, but the structure of large-scale flow with high correlation of vertical vorticity and vertical velocity is very promising for helicity formation.

\section{Experimental setup and numerical model}

\begin{figure}[ht!]
\begin{minipage}{0.45\linewidth}
\center{\includegraphics[width=0.86\linewidth]{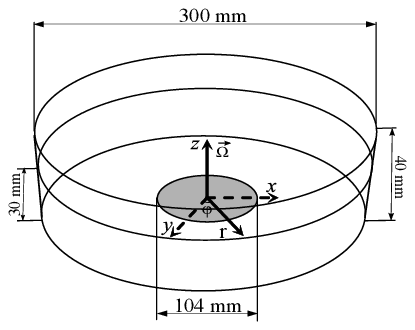}}\\
\end{minipage}
\hfill
\begin{minipage}{0.45\linewidth}
\center{\includegraphics[width=0.86\linewidth]{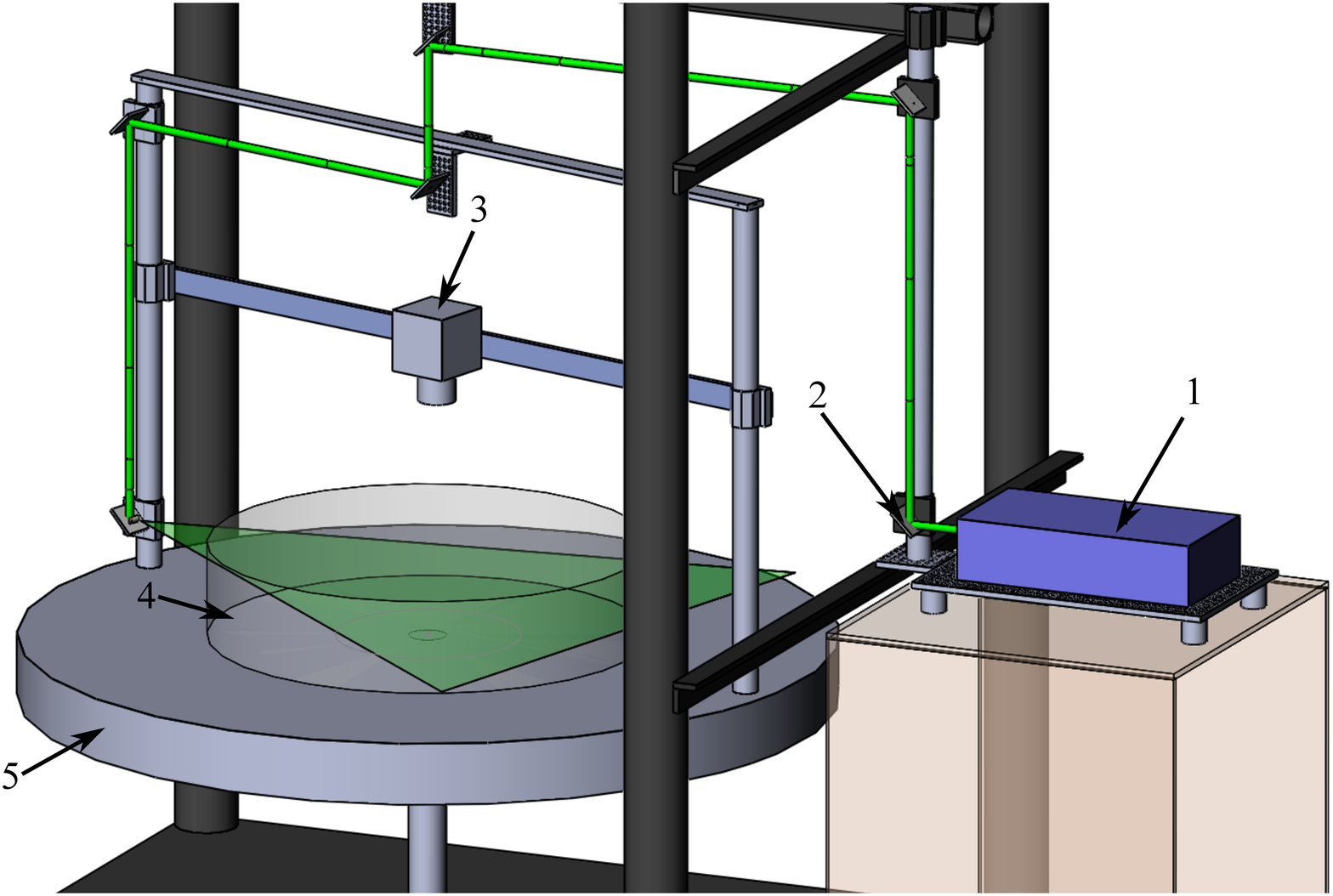}}\\
\end{minipage}
\caption{Experimental model, dimensions and location of the coordinate system. T - thermocouple for control of the mean temperature (left); Experimental setup: 1 - dual pulsed laser for PIV, 2 - laser sheet system, 3 - CCD camera, 4 - experimental model, 5 - rotating table (right). }
\label{Fig.1}
\end{figure}

Experimental model is a cylindrical vessel of diameter $D = 300$ mm, and height $L = 40$ mm (Fig.~\ref{Fig.1}). The sides and the bottom were made of Plexiglas with a thickness 3 mm and 20 mm respectively. There was no cover or additional heat insulation at the sidewalls. The heater is a brass cylindrical plate mounted flush with the bottom. The diameter of the plate $d$ is 104 mm, and its thickness is 10 mm. The brass plate is heated by an electrical coil placed on the lower side of the disc. Cylindrical vessel was placed on a rotating horizontal table (Fig.~\ref{Fig.1}). Silicon oil with  kinematic viscosity value of 5 sSt (T=25$^0$C) was used as working fluid. In all experiments, the depth of the fluid layer $l$ was 30 mm and the  surface of the fluid was always open. The room temperature was kept constant by air-conditioning system, and cooling of the fluid was provided mainly by the heat exchange with surrounding air on the free surface and some heat losses through sidewalls. For low values of kinematic viscosity it takes about 2 hours to obtain a steady-state regime. Temperature inside the fluid layer was measured at mid-height ($z = 15$ mm), near the periphery (about 3 cm from the sidewall)  by copper-constantan thermocouple. It was used for the estimation of the mean temperature of the fluid. The velocity field measurements were made with a 2D particle image velocimetry (PIV) system Polis. The system included a dual pulsed Nd-YaG laser, a control unit, a digital CCD camera (11 megapixels), placed in a rotating frame, and a computer. The synchronization of the operation of the laser and the CCD camera, the measurement, and the processing of the results were performed using the software package Actual Flow. Cylindrical vessel works as a lens and narrow horizontal light sheet from the periphery to the centre, but all area of our interest in the central part of the vessel was illuminated. Also, we need to note that, due to strong optical distortions, we did not make PIV measurements in a close proximity to the heater at height less than 2 mm. Iterative PIV algorithms [11] and decreasing of the size of the interrogation windows from $32\times32$ to $16\times16$ pixels provided a dynamic range of approximately 500 (the ratio of the maximum and minimum resolvable particle displacement). The PIV velocity measurements were accurate to within 5\%, estimated from calibration experiments in solid body rotation and long time series.

Along with the dimensional parameters (heating flux, rotation rate and kinematic viscosity) we use the set of the non-dimensional parameters which are commonly used for similar problems and can help for comparison our results with the results obtained by other researchers.  These are the flux Grashof number ${Gr_f}$, non-dimensional rotation velocity $Re$, Ekman number $E$ and Prandtl number $Pr$ :
\begin{equation}
 {{Gr_f}} = \frac{g \beta l^4 q }{ c \rho \kappa \nu^2 } \label{Grashof}
\end{equation}
\begin{equation}
{Re} =\frac{\Omega l^{2}}{\nu} \label{Reynolds}
\end{equation}
\begin{equation}
{E} =\frac{\nu}{\Omega l^{2}} \label{Ekman}
\end{equation}
\begin{equation}
 {{Pr}} = \frac{\nu}{ \kappa } \label{Prandtl}
\end{equation}
where $g$ is the gravitational acceleration, $l$ is the layer depth, $\beta$  is the coefficient of thermal expansion, $c$ is the thermal capacity, $\rho$ is the density, $\nu$ is the coefficient of kinematic viscosity and $\kappa$ is the thermal diffusivity, $q$ is a heat flux ($q=P/S_h$, $P$ is the power of the heater and $S_h$ is the heater's surface area). The value of non-dimensional rotation velocity $Re$ is equal to the inverse value of Ekman number $E$, but because of different physical meaning of these parameters it is convenient to use them both.

For the analysis of helicity distribution it is necessary to have 3D velocity field. There are some experimental techniques that provide 3D velocity measurements, for example tomographic particle image velocimetry (Tomo-PIV), but they are very expensive and resource-demanding. So, up to now numerical simulations are the main tool for studying complex three-dimensional flows.  In our case we used experimental data only for verification of the numerical model. General results were obtained using the CFD package FlowVision. All numerical runs were done in 3D statement and the integration domain was a cylindrical cavity, similar to the one used in the laboratory experiments. The numerical finite volume code is used to solve the Boussinesq equations for thermal convection. The fluid is assumed to be Newtonian and the flow is considered to be incompressible and laminar. We used the implicit splitting scheme of second order of accuracy. Impermeable and no-slip velocity conditions were applied at the side wall and bottom. The upper boundary was stress-free. The bottom had a localized heat source in the central part defined through a heat flux and the diameter of the heating area was fixed at D = 100 mm. The upper surface was cooled by uniform heat flux. The net heat flux (heating and cooling) was zero. Spatial resolution was 1 mm in all directions. Time step was 0.1 s, the time of calculation was about 600 seconds. Physical properties of the working fluid were chosen similar to the experiment. The depth of the fluid layer was 30 mm. In Table 1 values of non-dimensional parameters for experiment and numerical simulation are showed.

\begin{table}[ht!]
\begin{center}
\caption{\label{tab5}Values of non-dimensional parameters for experiment and numerical simulation}
\begin{tabular}{|c||c||c||c||c|}
&$Gr_f \cdot 10^6$&Re&E&Pr\\
\hline
FlowVision&4.6&29&0.035&61\\
\hline
experiment&4.5&27&0.037&60\\
\end{tabular}
\end{center}
\end{table}

We performed a mesh sensitivity analysis for our calculations. We used 6 models with different mesh resolution and time discretization. The change of time step from 0.05 s to 0.1 s had very weak influence on numerical results. Mesh resolution is a more important parameter. In table \ref{tab6}, time averaged volume integrals of kinetic energy are presented for different mesh resolutions. Values of volume kinetic energy for mesh 1 mm and 0.5 mm are very close in comparison with mesh 2 mm so we decided that the mesh 1 mm and time step 0.1 s is adequate  choice for numerical simulation with limited computing resources.

\begin{table}[ht!]
\begin{center}
\caption{\label{tab6} Time average integral values of kinetic energy}
\begin{tabular}{|c||c||c||c|}
mesh&0.5 mm&1mm&2mm\\
\hline
$E_k\cdot 10^{-6}, m^5/s$&0.0243&0.0244&0.0233\\
\end{tabular}
\end{center}
\end{table}

\begin{figure}[ht!]
\begin{minipage} {0.48\linewidth}
\center{\includegraphics[width=1\linewidth]{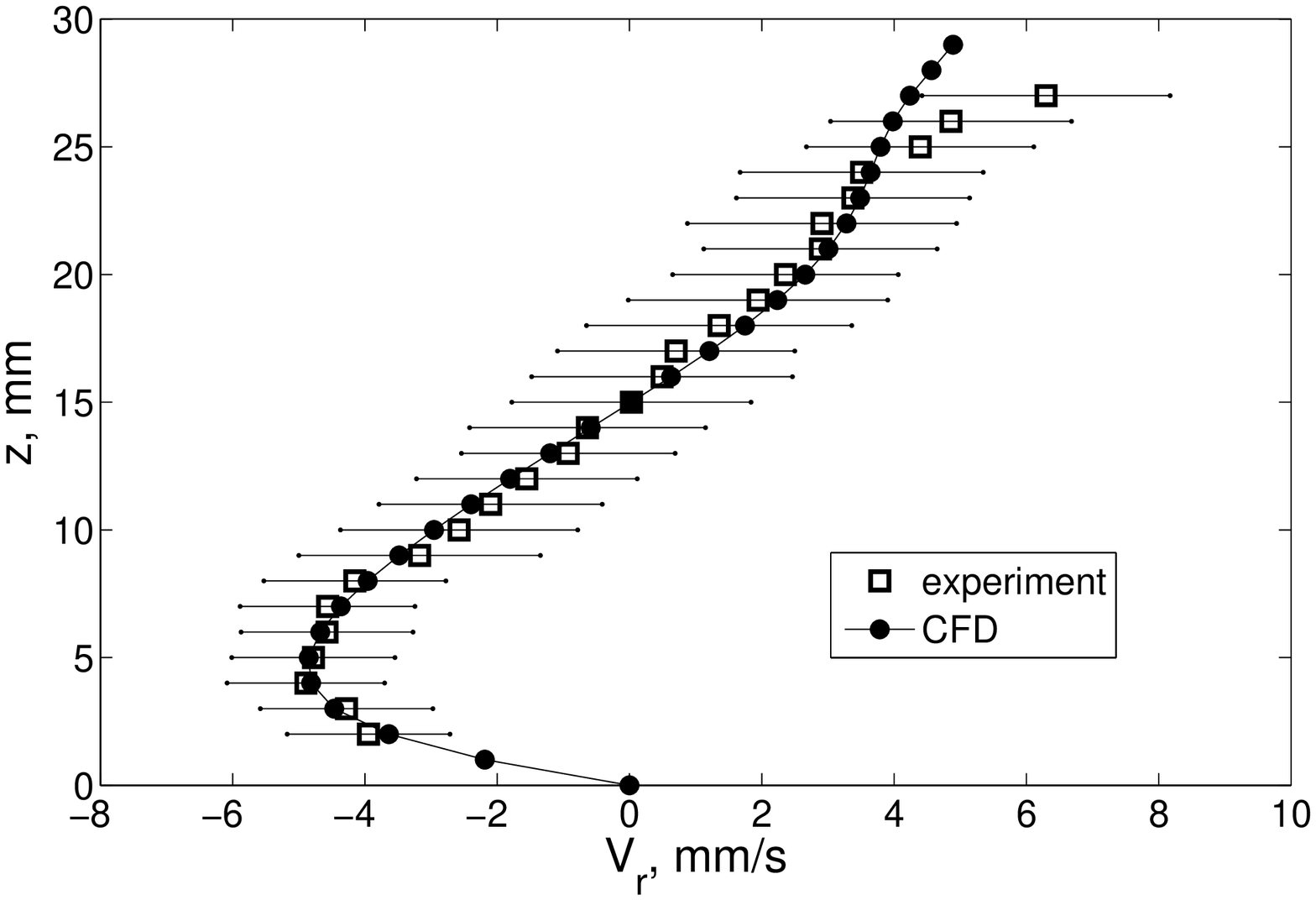}
\\ a}
\end{minipage}
\hfill
\begin{minipage} {0.48\linewidth}
\center{\includegraphics[width=1.05\linewidth]{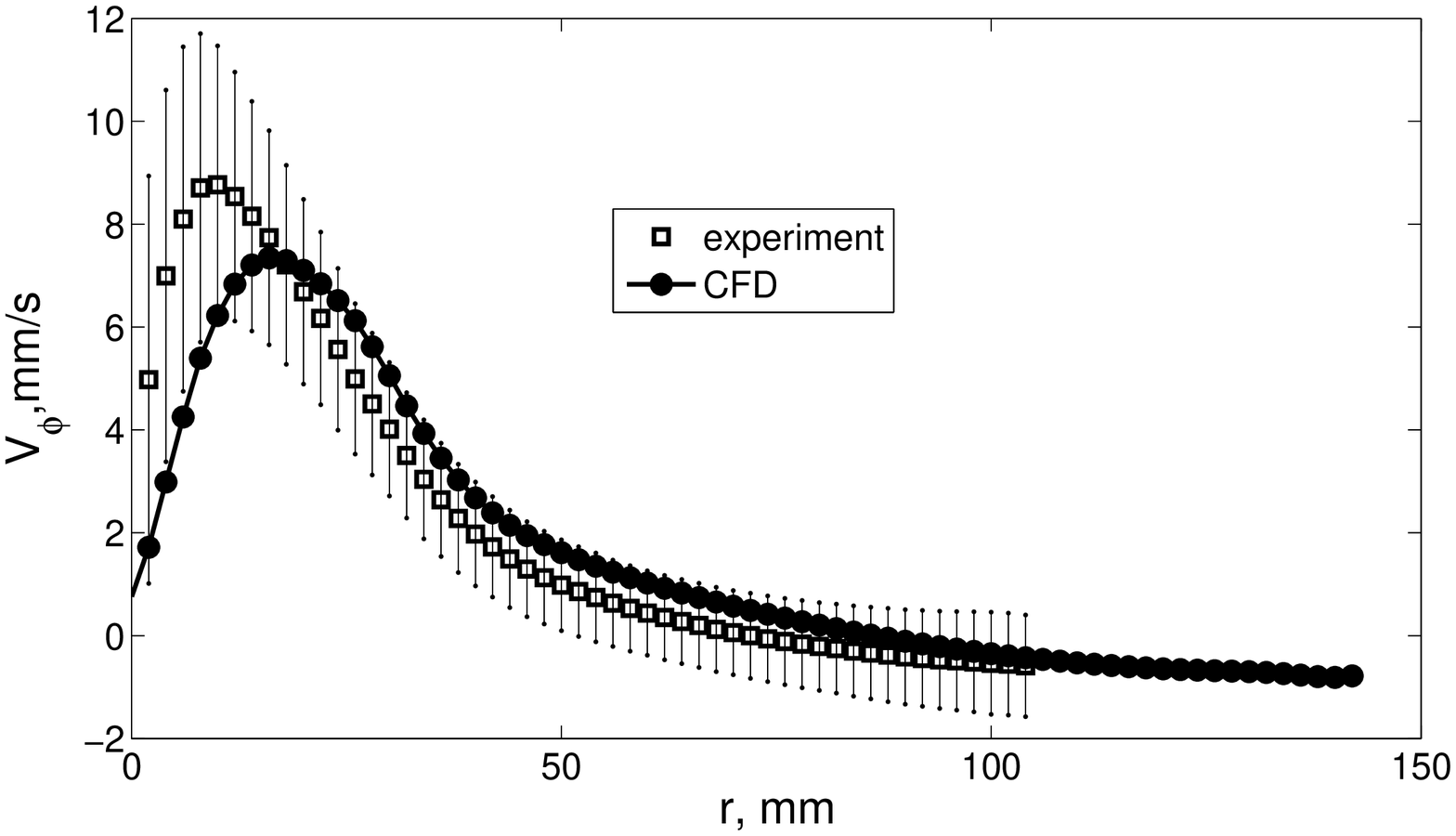}
\\ b}
\end{minipage}
\caption{Verification of numerical model: a- vertical profile of mean radial velocity at r=15mm, b-horizontal profile of mean azimutal velocity at z=15mm}
\label{VrVfprof}
\end{figure}

For verification of numerical results, we compared vertical profiles of the mean radial velocity at r=15 mm and horizontal profiles of azimuthal velocity at z=15 mm obtained from experimental measurements with rms (root mean square) and numerical calculations (Fig.~\ref{VrVfprof}). Velocities are averaged in the azimuthal direction and in time. One can see that the mean velocity profiles are in a good agreement.

\section{General structure of the flow}

\begin{figure}[ht!]
\begin{center}
\includegraphics[width=.4\textwidth]{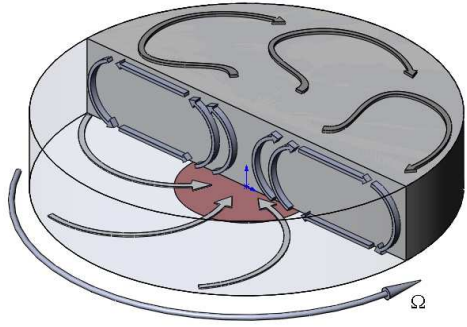}
\caption{Scheme of large-scale circulation } \label{Fig.2}
\end{center}
\end{figure}

The heat flux in the central part of the bottom initiates the intensive upward motion above the heater. Warm fluid cools at the free surface and moves toward the periphery where the cooled fluid moves downward along the side wall. After some time, large-scale advective flow occupies the whole vessel (Fig.~\ref{Fig.2}, vertical cross-section). Experimental measurements of velocity fields in a non-rotating layer, in a vertical cross-section over the heating area showed that instantaneous fields are irregular and asymmetric. Along with the main updraft in the centre there are less intensive but pronounced upgoing convective flows close to the periphery of the heater.

\begin{figure}
\begin{minipage} {0.45\linewidth}
\center{\includegraphics[width=1\linewidth]{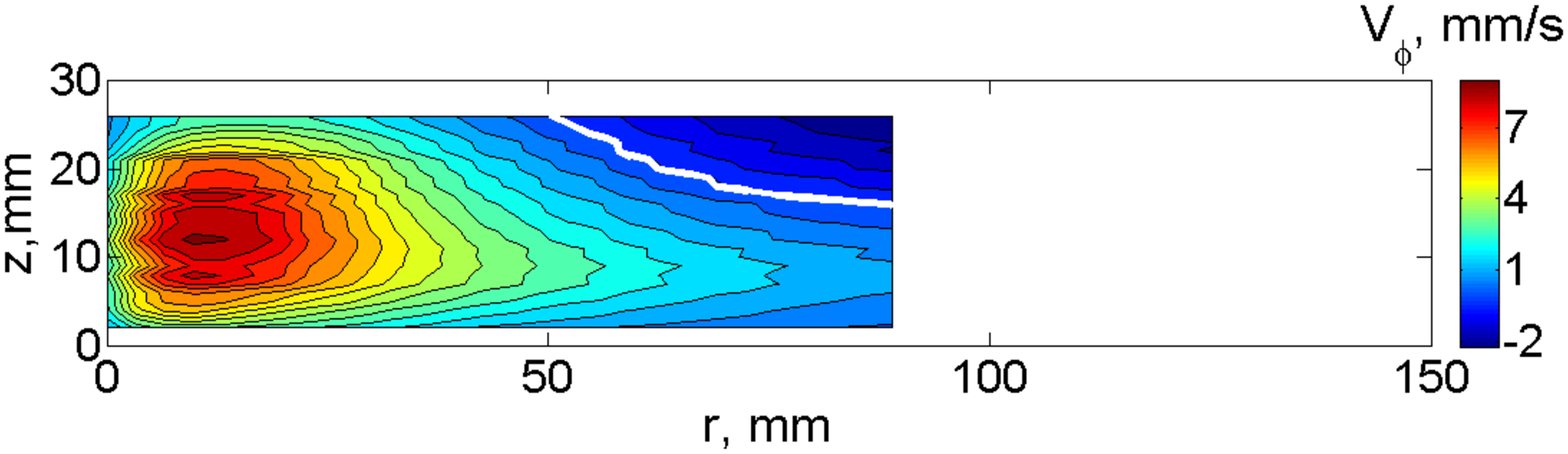}
\\ a}
\end{minipage}
\hfill
\begin{minipage} {0.45\linewidth}
\center{\includegraphics[width=1\linewidth]{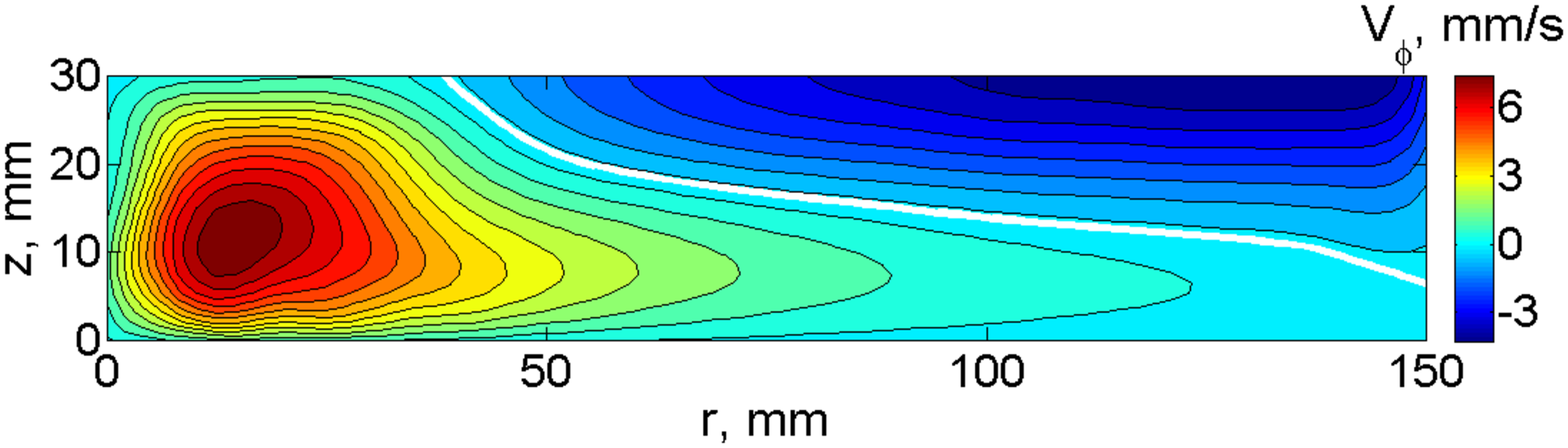}
\\ b}
\end{minipage}
\caption{Mean azimuthal velocity fields for: (a) experiment, (b) FlowVision. Positive values describe cyclonic motion, negative - anticyclonic motion, thick solid white line shows the border between cyclonic and anticyclonic flows.}
\label{Fig.5}
\end{figure}

The structures of the steady-state azimuthal flows (in a rotating frame) for numerical model and experimental case are shown in  Fig.~\ref{Fig.5}. Positive (negative) values of velocity describe cyclonic (anticyclonic) motion.  Distribution of azimuthal velocity is qualitatively similar to the one of a mature tropical cyclone [8]. The cyclonic vortex formation in the laboratory system can be described by the following scenario. Large-scale radial circulation leads to the angular momentum transport and the angular momentum exchange on the solid boundaries. Convergent flow in the lower layer brings the fluid parcels with large values of angular momentum from the periphery to the centre and produces cyclonic motion (Fig.~\ref{Fig.2}, lower horizontal cross-section). In the upper layer situation is the opposite - divergent flow takes the fluid with low values of angular momentum to the periphery resulting in anticyclonic motion (Fig.~\ref{Fig.2}, upper horizontal cross-section).  Friction in the viscous boundary layers leads to the sink of angular momentum in the part of the bottom occupied by cyclonic flow and produces a source of angular momentum on the sidewalls when anticyclonic flow comes to the periphery. Zero net angular momentum flux on the solid boundaries is the necessary condition  for the steady-state regime. In [5] it was shown that steady-state cyclonic vortex can exists only in a certain range of dimensional parameters, relatively small variations of viscosity or rotation rate may remarkably change the cyclonic vortex structure and its intensity. Here we consider interval of governing parameters, which provides existence of steady convective cyclonic vortex in the central part of domain.

\begin{figure}
\begin{minipage} {0.45\linewidth}
\center{\includegraphics[width=0.8\linewidth]{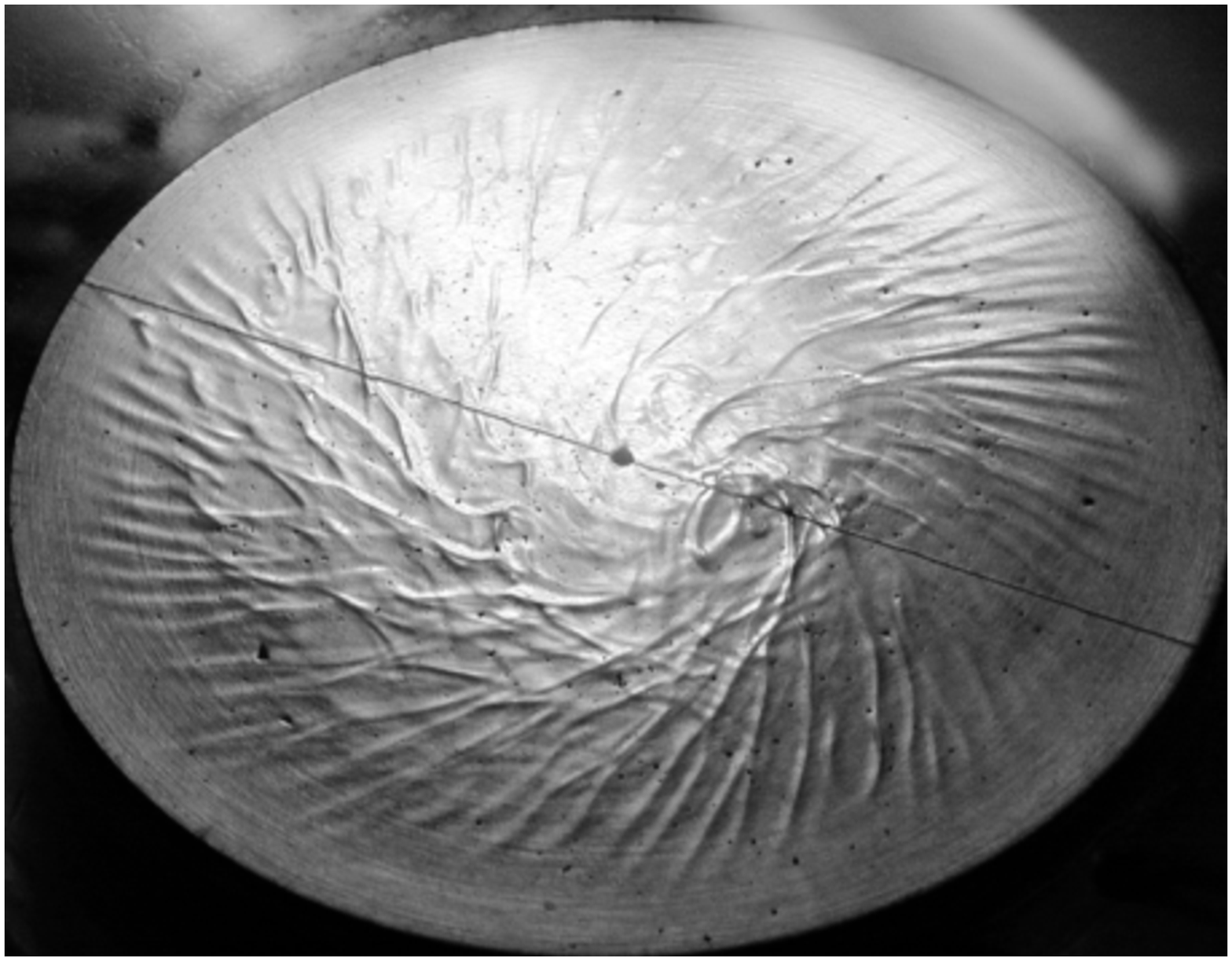}
\\ a}
\end{minipage}
\hfill
\begin{minipage} {0.45\linewidth}
\center{\includegraphics[width=1\linewidth]{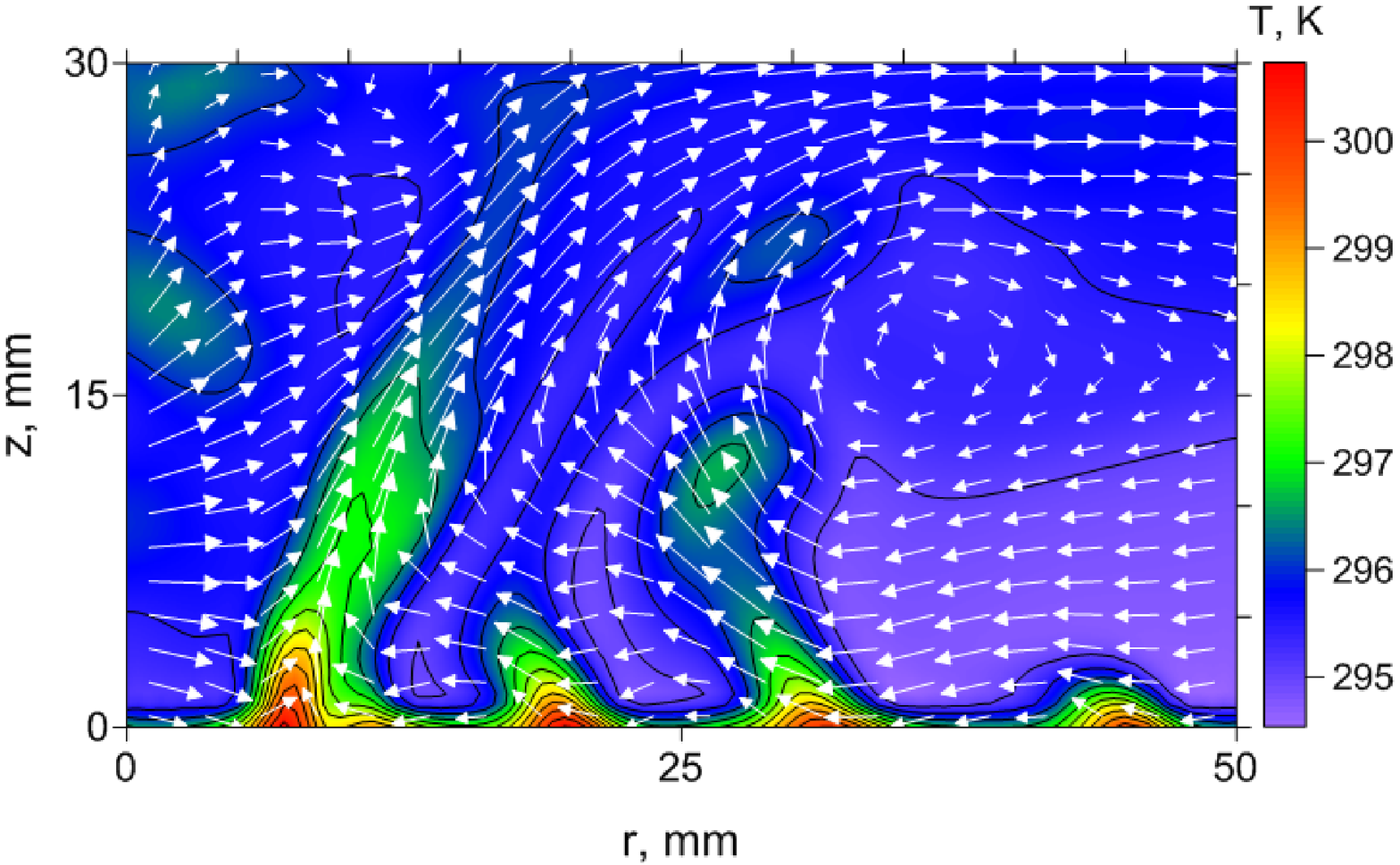}
\\ b}
\end{minipage}
\begin{minipage} {0.45\linewidth}
\center{\includegraphics[width=1\linewidth]{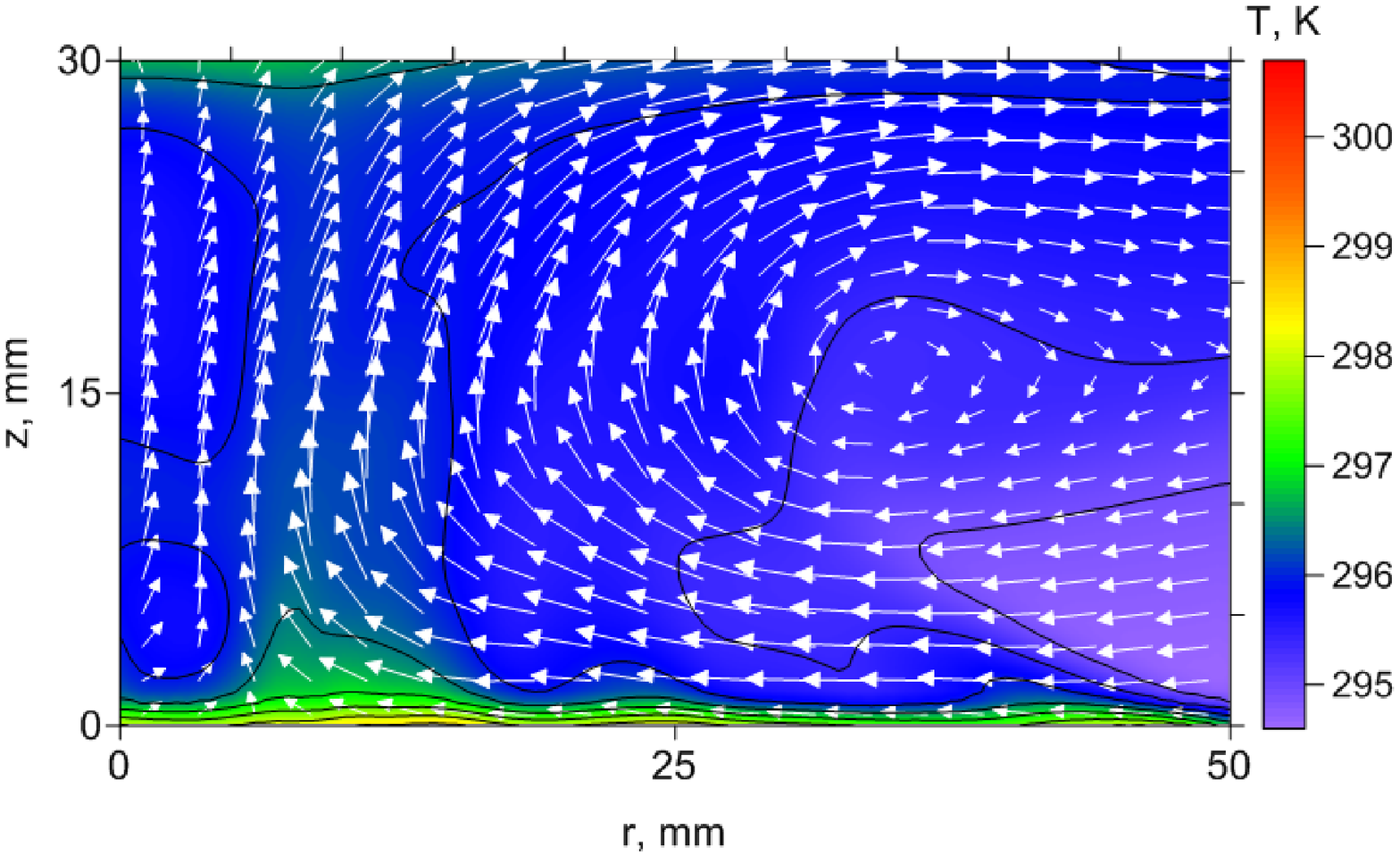}
\\ c}
\end{minipage}
\caption{a - Shadowgraph image of the secondary flows in the thermal boundary layer over the heater $Gr_f = 1.7\cdot10^7, Re = 30$; b, c - instantaneous and mean temperature and velocity fields in vertical cross-section obtained by FlowVision $Gr_f = 4.5\cdot10^6, Re = 27$}
\label{Fig.6}
\end{figure}

The large-scale advective flow in the lower part of the layer leads to the formation of boundary layer with potentially unstable temperature stratification above the heater and makes possible the generation of the secondary convective flows. The structure and specifics of secondary flows over the heater in the case of non-rotating cylindrical layer are described in detail in [12]. Regimes for the weak heating are characterized by appearance of ring-like rolls. These rolls are permanently shifted to the centre by the main flow. The increase in heating leads to the instability of convective transverse rolls and, instead of transverse rolls, there appear radial rolls. Various types of visualization revealed co-existence of convective radial rolls and transverse rolls of another kind. In [12] was proposed the following scenario of secondary structures formation. Neighbouring radial rolls produce alternating areas with positive and negative values of vorticity. One pair of rolls brings relatively cold fluid down and another pair takes warm fluid up. Periodically overheated parcels of fluid continue to float and produce thermal plumes moving to the centre with the main flow. A convergent flow in the lower layer leads to the merging of thermal plumes and the formation of a transverse convective roll of a different type. Visualization of the secondary structures over the heater in the rotating case with shadowgraph method is shown in Fig.~\ref{Fig.6}. Thermal plumes are clearly seen in the instantaneous  temperature field in a vertical cross-section obtained by numerical simulation (Fig.~\ref{Fig.6}), time average temperature and velocity fields are showed in Fig.~\ref{Fig.6}c . It should be noted that secondary flows disturb velocity field and thereby fields of vorticity, so secondary structures are also important for the process of helicity formation.

The flow in the proposed configuration is very complex and consists of different scale structures. For better understanding of helicity distribution we divided it into three parts - radial, azimuthal and vertical helicity (\ref{helphirz}).

\begin{equation}
h_{\phi} = V_{\phi}\cdot \omega_{\phi};\qquad h_r = V_r \cdot \omega_{r};\qquad h_z = V_z \cdot \omega_{z}
\label{helphirz}
\end{equation}
where $\omega_{\phi}, \omega_{r}, \omega_{z}$ - azimutal, radial and vertical components of vorticity (\ref{wrphiz}).
\begin{equation}
\omega_{\phi}=-\frac {\partial V_z}{\partial r}+\frac {\partial V_r}{\partial z}; \qquad \omega_r=\frac1{r}\frac{\partial V_{z}}{\partial \phi}-\frac{\partial V_{\phi}}{\partial z};\qquad \omega_{z}=\frac1 {r}\frac {\partial r V_{\phi}}{\partial r}-\frac1{r} \frac {\partial V_r}{\partial \phi}
\label{wrphiz}
\end{equation}

\section{Helicity in rotation fluid layer with localized heat source}

It is common in fluid dynamics and the theory of turbulence to separate the average and fluctuating parts. For example, for a velocity the decomposition would be (\ref{pulVW}) as well as for vorticity.

\begin{equation}
\upsilon(r, \phi, z, t)=\bar \upsilon (r, \phi, z)+\upsilon'(r, \phi, z, t) \qquad \omega(r, \phi, z, t)=\bar \omega(r, \phi, z, t)+\omega'(r, \phi, z, t)
\label{pulVW}
\end{equation}

where $\bar \upsilon, \bar \omega$ denotes the time average of $\upsilon$ and $\omega$ (often called the steady component), and $\upsilon', \omega'$, the fluctuating part (or perturbations). The time average of perturbations equals zero.

For helicity, this decomposition will be the same, but the mean component of helicity $\bar h$ would include time average product of velocity and vorticity perturbations $<\upsilon' \omega'>_t$ (\ref{pulH}).

\begin{equation}
h(r, \phi, z, t) = \bar h(r, \phi, z)+h'(r, \phi, z, t); \qquad \bar h= \bar \upsilon \cdot \bar \omega+ <\upsilon' \omega'>_t;
\label{pulH}
\end{equation}

\begin{figure}[ht!]
\center{\includegraphics[width=0.6\linewidth]{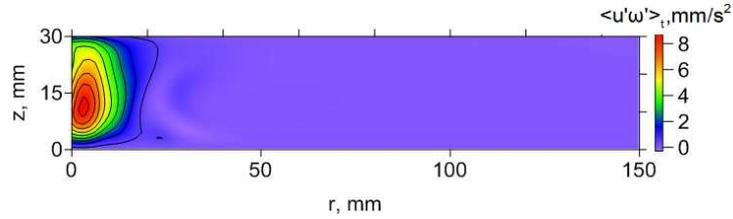}}
\caption{The time and azimuthal average of fluctuating part of mean helicity}
\label{VpRp}
\end{figure}

Fluctuating part of mean helicity $<\upsilon' \omega'>_t$, unlike velocity and vorticity time averages ($\bar {\upsilon'}, \bar {\omega'}$), is not zero. In our case the fluctuating part is approximately 12 \% of mean helicity. The average (on time and azimuthal coordinate) field of $<\upsilon' \omega'>_t$ is presented in Fig.\ref{VpRp}. The high correlation of velocity and vorticity perturbations is located in the area of intensive upward flow above the heater.

\begin{figure}[ht!]
\center{\includegraphics[width=0.9\linewidth]{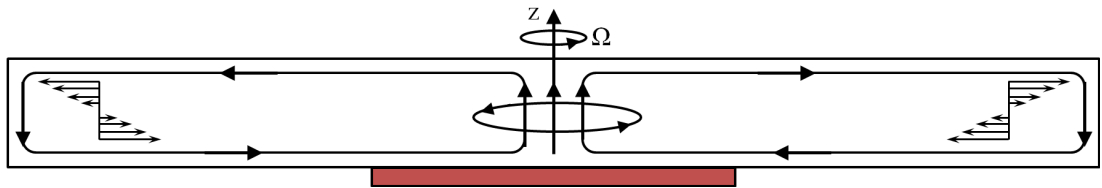}}
\caption{Mean flow structure in a vertical cross-section.}
\label{flowstructurerot}
\end{figure}

In further discussion, we will analyze the mean helicity ($\bar h \equiv h$) including fluctuations.
There are two main mechanisms that can lead to the existence of helicity in a described system. Fig.~\ref{flowstructurerot} shows mean flow structure in a vertical cross-section. As a first mechanism we assume strong correlation between upward flow and vertical vorticity in a central area, over the heater. As a second mechanism we consider strong shear of radial and azimuthal velocities on the periphery, which also produces helicity.

\begin{figure}[ht!]
\begin{minipage} {0.48\linewidth}
\center{\includegraphics[width=1\linewidth]{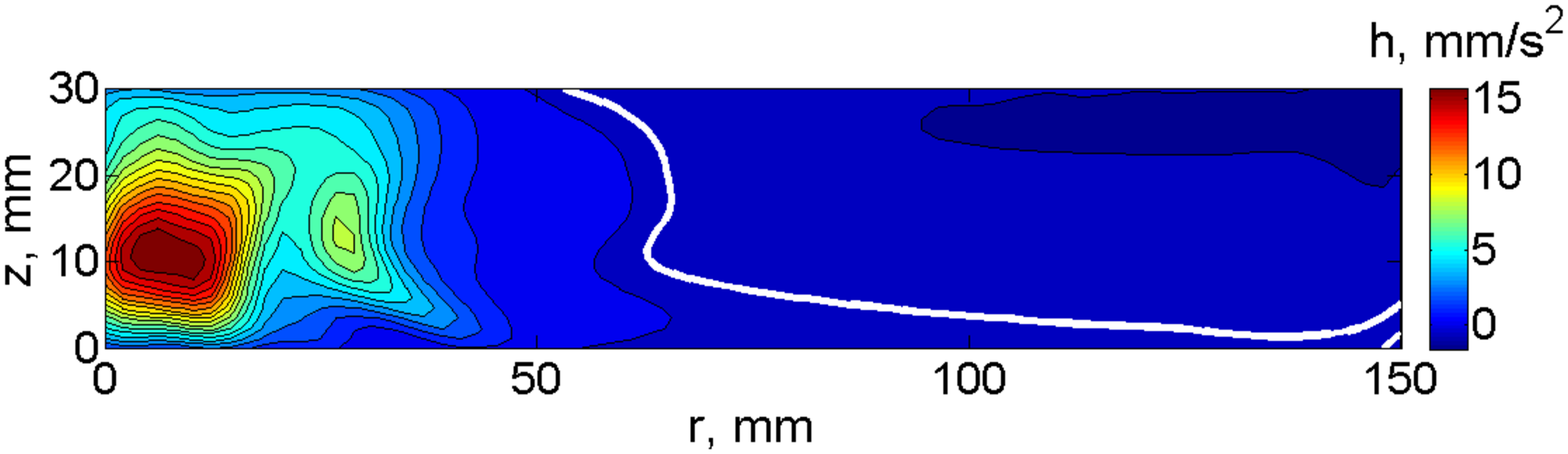}
\\ a}
\end{minipage}
\hfill
\begin{minipage} {0.48\linewidth}
\center{\includegraphics[width=1\linewidth]{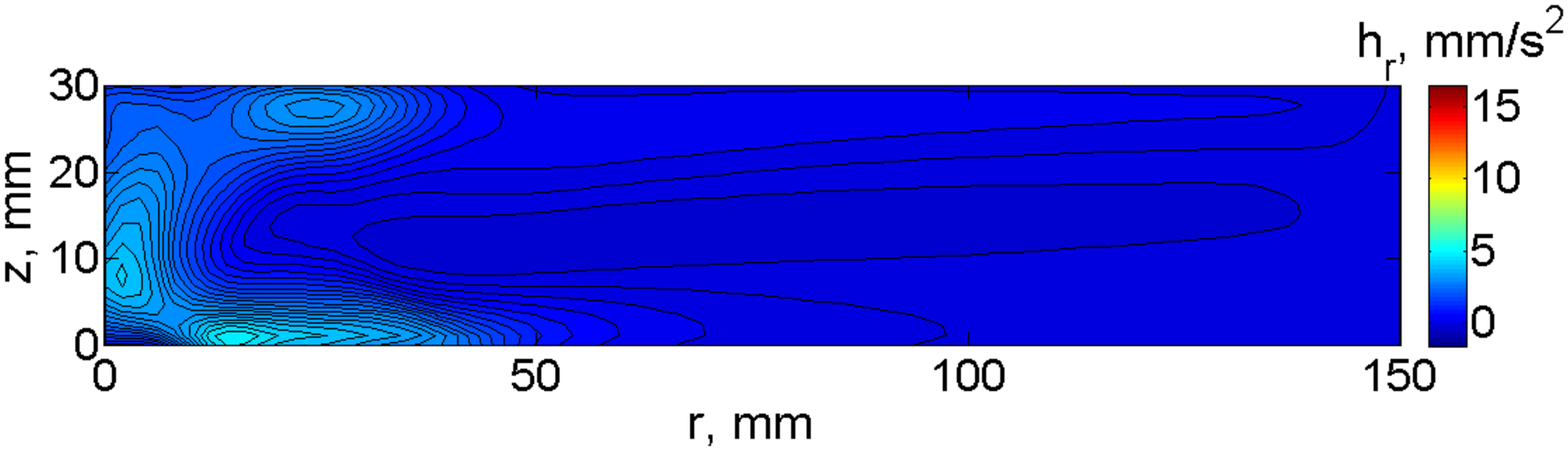}
\\ b}
\end{minipage}
\vfill
\begin{minipage} {0.48\linewidth}
\center{\includegraphics[width=1\linewidth]{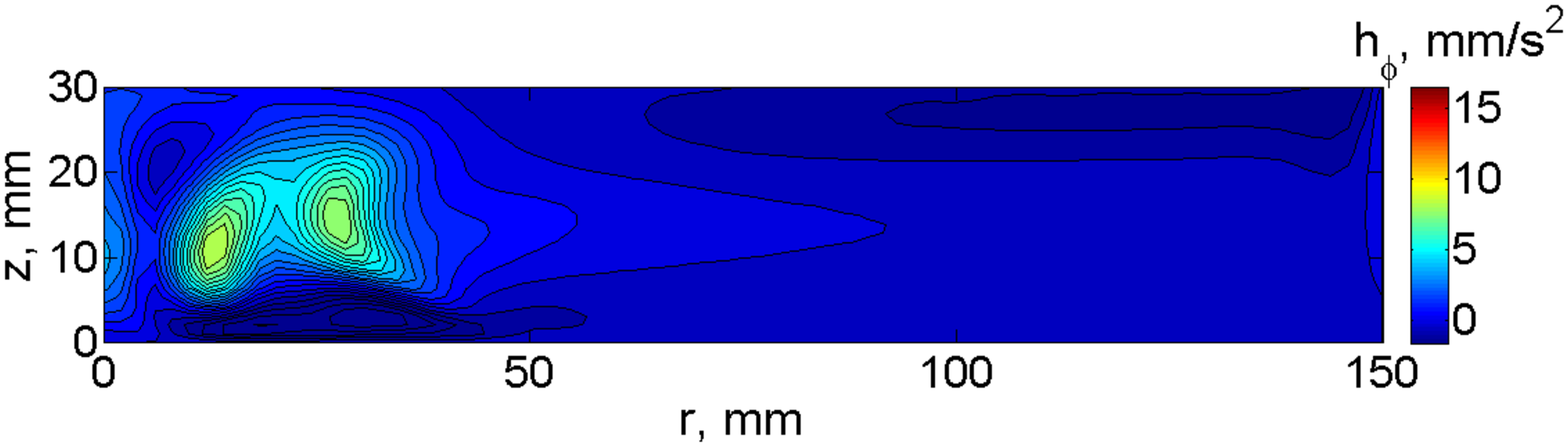}
\\ c}
\end{minipage}
\hfill
\begin{minipage} {0.48\linewidth}
\center{\includegraphics[width=1\linewidth]{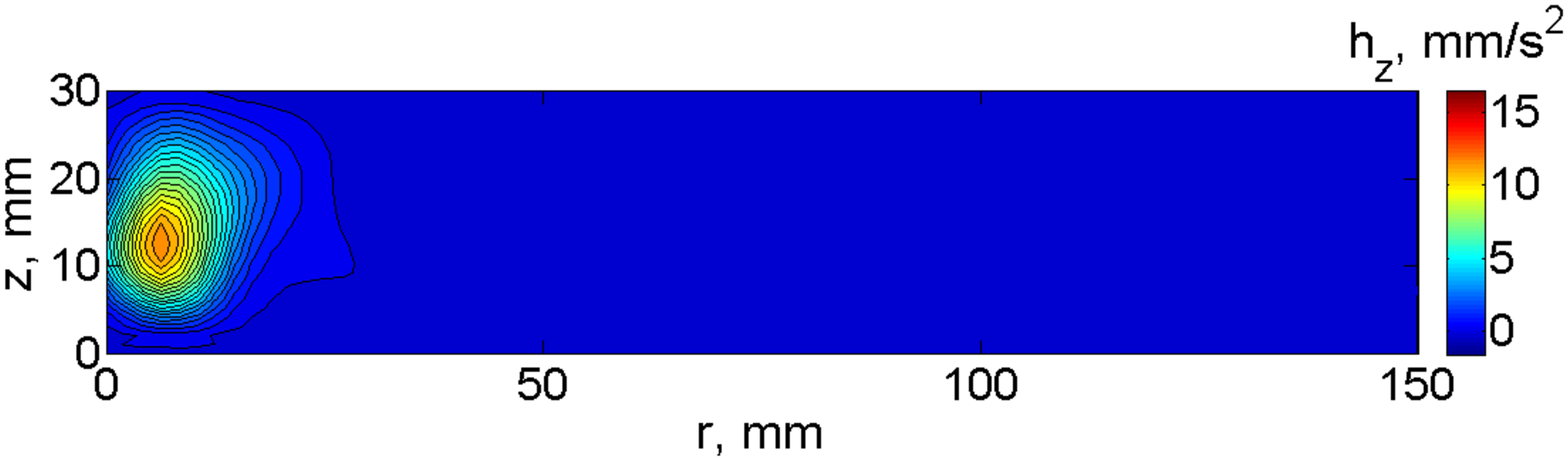}
\\ d}
\end{minipage}
\caption{Gr$_f$ =$4.6\cdot10^6$: a- mean helicity, helicity components - radial (b), azimuthal (c) and vertical (d)}
\label{Hsrpoa}
\end{figure}

Fig.~\ref{Hsrpoa} shows mean helicity in the vertical cross-section (a), their components are presented in Fig.~\ref{Hsrpoa} b, c, d. As we have assumed, there is a dominance of positive helicity in the central part. Besides vertical helicity due to existence of intensive cyclonic vortex, there are significantly large values of azimuthal helicity. Azimuthal helicity can appear because of the small-scale convective plumes. Convective plumes disturb velocity fields and lead to the large gradient of vertical velocity in radial direction. In addition to the cyclonic motion in the central part, it provides azimuthal helicity generation.

\begin{figure}[ht!]
\begin{minipage} {0.48\linewidth}
\center{\includegraphics[width=1\linewidth]{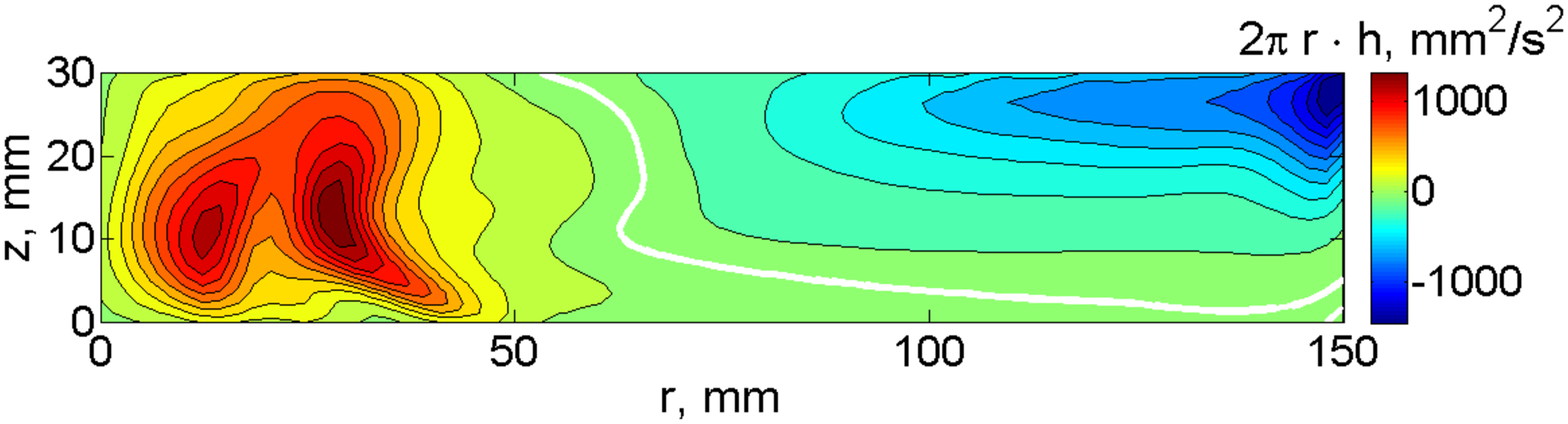}
\\ a}
\end{minipage}
\hfill
\begin{minipage} {0.48\linewidth}
\center{\includegraphics[width=1\linewidth]{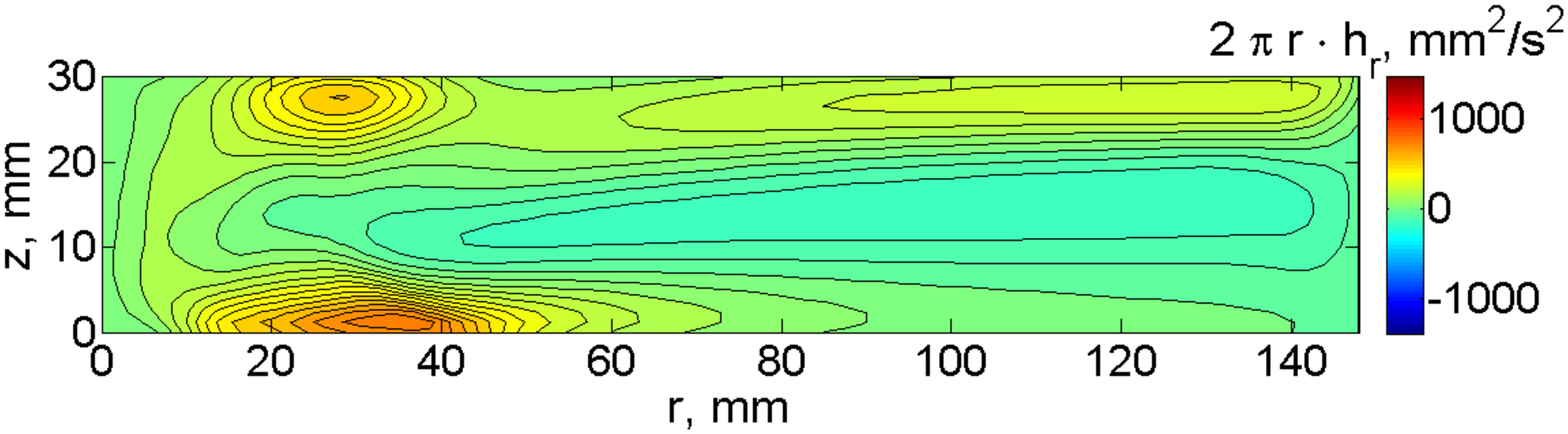}
\\ b}
\end{minipage}
\vfill
\begin{minipage} {0.48\linewidth}
\center{\includegraphics[width=1\linewidth]{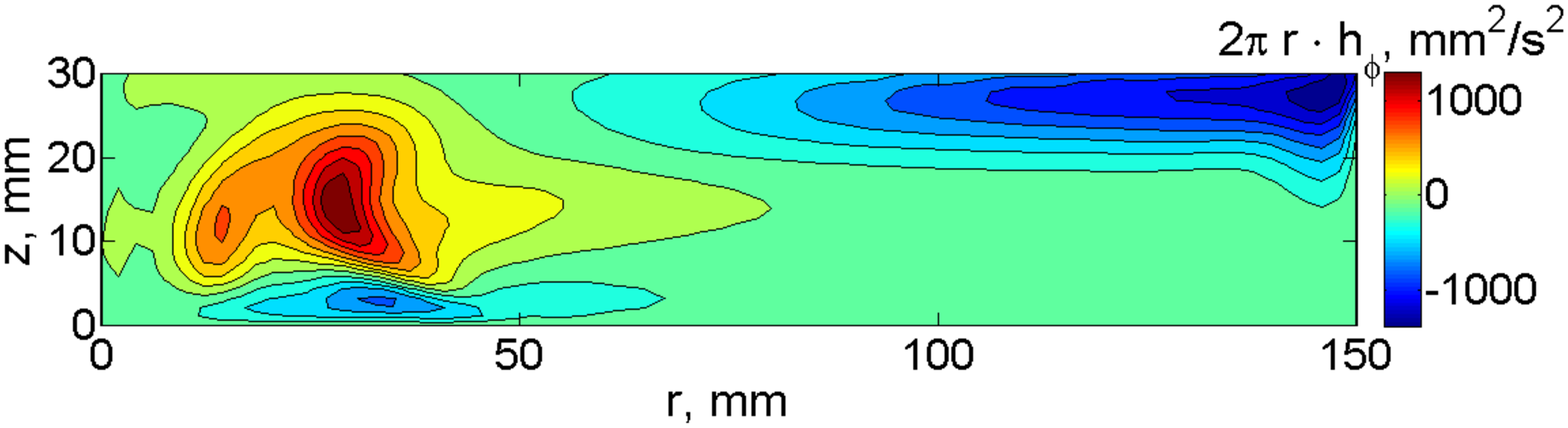}
\\ c}
\end{minipage}
\hfill
\begin{minipage} {0.48\linewidth}
\center{\includegraphics[width=1\linewidth]{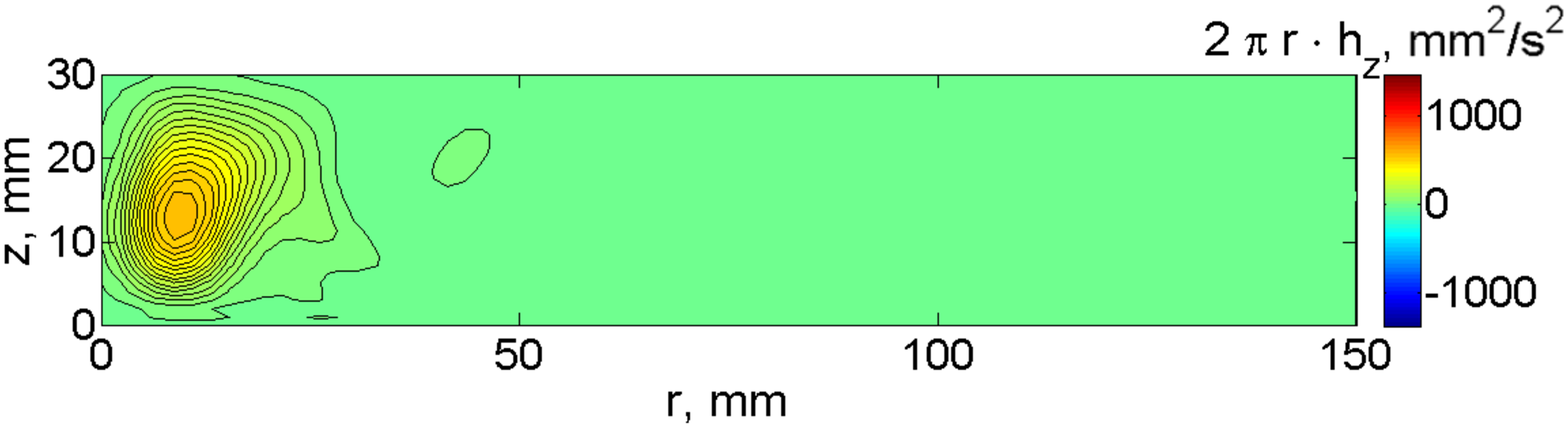}
\\ d}
\end{minipage}
\caption{Gr$_f$ =$4.6\cdot10^6$: a- mean helicity, helicity components - radial (b), azimuthal (c) and vertical (d)}
\label{intHsrpoa}
\end{figure}

In addition to analysis of helicity distribution in a vertical cross-section, it is very important to study integral (over azimuthal coordinate) fields, because even weak local helicity in the periphery may results in substantial values after integration. In Fig.~\ref{intHsrpoa}, azimuthally integrated helicity and each of their component are presented. After integration, we have found substantial values of negative azimuthal helicity in the periphery in the upper layer. Negative values are located in the area of  large gradients of radial velocity, when the divergent flow is replaced by a convergent one. Interacting with anticyclonic motion, it provides negative values of azimuthal helicity. Positive values of global helicity are in the central area, where cyclonic vortex and convective plumes play a major role. Asymmetrical distribution of integral helicity provides non-zero total value. This is an important result, because it proves that the flow in the described system is characterized by substantial values of helicity.

The relative level of helicity is estimated by the dimensionless ratio $|H(k)|/2kE(k)$, where $k$ is the wave number and $H(k), E(k)$ - spectral densities of helicity and kinetic energy [13]. This ratio is usually used to estimate the influence of helicity on spectral properties of turbulent flows [14] and references therein]. In our case, for estimation of the helicity level we multiply integral helicity ($H$) by characteristic size (depth of the layer) in order to have the same dimension with integral kinetic energy (E) and tu calculate their ratio (Tab.\ref{tab9}). In our case, the relative level of helicity $H\cdot l / 2E$ is approximately 20 $\%$. In the non-rotating layer this ratio was less than 1 $\%$ ([15]).

\begin{table}[ht!]
\begin{center}
\caption{\label{tab9}}
\begin{tabular}{|c||c||c|}
$H, m^4/s^2$ &$E, m^5/s^2$&$H\cdot l / 2E$\\
\hline
0.68$\cdot 10^{-8}$&0.57$\cdot 10^{-9}$& 0.18\\
\end{tabular}
\end{center}
\end{table}

\section{Conclusion}
In this paper, we analyzed helicity distribution in a laboratory model of a tropical cyclone. The vortex is formed in a rotating fluid layer with a localized heat source. Two mechanisms which play role in helicity generation are defined. The first one is the strong correlation of cyclonic vortex and intensive upward motion in the central part of the vessel. The second one is due to large gradients of velocity on the periphery. It was shown that, besides these mechanisms, there is one more factor that can be a crucial part in helicity generation. It is a system of secondary flows appearing over the heating area. Thermal plumes appearing over the heater disturb the field of velocity and vorticity. It leads to existing high values of integral helicity in the place of convective plumes ascending. The integral helicity in the considered case is substantial and relative level of helicity is high. It proves that the chosen configuration is very promising and requires further detailed studies for a wide range of governing parameters.

\section{Acknowledgments}
The author wants to express gratitude to the organizers of 8th European Postgraduate Fluid Dynamics Conference for possibility of presenting partial results published in this paper.

We sincerely thank Peter Frick for constructive comments which led to serious improvement of the article.

This work was supported by the Russian Science Foundation (grant No. 16-41-02012).

\bibliographystyle{iopart-num}

\section*{References}

\numrefs{99}
\item Bogatyrev GP 1990 {\it Pisma Zh.Eksp. Teor. Fiz.} {\bf 51} 557-559.
\item Bogatyrev GP and Smorodin BL 1996 {\it Pisma Zh.Eksp. Teor. Fiz.} {\bf 63} 25-28.
\item Bogatyrev GP et al. 2006 {\it Izvestiya Atmospheric and Oceanic Physics} {\bf 42} 423-429.
\item Batalov et al.  2010 {\it Geophys. Astrophys. Fluid Dyn.} {\bf 104} 349-368.
\item Sukhanovskii et al. 2016 {\it Quart J.R.Met.Soc.} {\bf 316} 23-33.
\item Moiseev et al. 1989 {\it Sov. Phys. Dokl.} {\bf 28} 926.
\item Levich E. and Tzvetkov E. 1984 {\it Physics Letters A} {\bf 100} 53-56.
\item Levich E. and Tzvetkov E. 1985 {\it Physics Reports} {\bf 128} 1-37.
\item Lilly D.K. 1986 {\it Journal of Atmospheric Sciences} {\bf 43} 126-140.
\item Eidelman et al. 2014 {\it Phys. Fluids} {\bf\ 26} 1-14.
\item Scarano and Riethmuller 2000 {\it Exp. Fluids} {\bf 29} 51-60.
\item Sukhanovskii et al. 2016 {\it Physica D} {\bf 316} 23-33.
\item Moffat H.K. 1978 {\it Cambridge University Press, Cambridge}.
\item Stepanov et al. 2015 {\it Physical Review Letters} {\bf 100} 1-5.
\item Sukhanovskii et al. 2013 {\it Computational Continuum mechanics} {\bf 6} 451-459.

\endnumrefs

\end{document}